# Band alignment of grafted diamond/GaN p-n heterojunctions interfaced with ALD $Al_2O_3$ and $SiN_x/Al_2O_3$

Tsung-Han Tsai[1], Chenyu Wang[1], Jiarui Gong[2], Xuanyu Zhou[3], Luke Suter[4], Aaron Hardy[4], Carolina Adamo[5], Yang Liu[1], Dong Liu[1], Connor S. Bailey[5], Michael Eller[5], Stephanie Liu[5], Matthias Muehle[4], Jung-Hun Seo[3,a)], Katherine Fountaine[5], Vincent Gambin[5], and Zhenqiang Ma[1,a)]

[1]*Department of Electrical and Computer Engineering, University of Wisconsin-Madison, Madison, Wisconsin, 53706, USA*

[2]*Department of Electrical and Computer Engineering, Texas A&M University, College Station, TX 77843, USA*

[3]*Department of Materials Design and Innovation, University at Buffalo, The State University of New York, Buffalo, NY 14260, USA*

[4]*Fraunhofer USA Inc, Center Midwest East Lansing, MI 48824, USA*

[5]*Northrop Grumman Corporation, Redondo Beach, CA 90278, USA*

[a)] These authors contributed equally to this work.

[b)] Author to whom correspondence should be addressed. Electronic mail: mazq@engr.wisc.edu or vincent.gambin@ngc.com, or katherine.fountaine@ngc.com

# ABSTRACT

Diamond and gallium nitride are complementary semiconductors for forming p–n junctions because of their respective doping limitations. Understanding the band alignment of grafted diamond/GaN heterojunctions is therefore essential for optimizing diode performance. In this study, the band alignment of diamond/$Al_2O_3$/GaN and diamond/$Al_2O_3$/$SiN_x$/GaN heterostructures was determined by X-ray photoelectron spectroscopy. Both structures exhibit type-II band alignment, but with different band offsets. The band offsets of the diamond/$Al_2O_3$/$SiN_x$/GaN heterojunction are larger by 0.42 eV than those of diamond/$Al_2O_3$/GaN. This difference is attributed to a modification of the interfacial electrostatic potential, which may arise from a reduced density of positive fixed charges in the interfacial dielectric near the diamond/$Al_2O_3$ interface after insertion of the $SiN_x$ layer. These results demonstrate that interfacial-layer engineering provides an effective strategy for tailoring the band alignment of grafted diamond/GaN heterojunctions, offering guidance for the design of p–n diodes with tunable rectifying characteristics.

## I. INTRODUCTION

Ultrawide-bandgap (UWBG) semiconductors are promising for future RF and power electronics.[1-3] Among them, diamond exhibits the highest figure of merit for power devices.[3-6] However, efficient formation of high-quality n-type diamond under non-extreme doping conditions remains elusive, making diamond p–n homojunctions extremely challenging.[7-10] Heterojunctions integrating diamond with n-type gallium oxide have been explored through epitaxial growth, direct bonding, and atomic layer deposition (ALD).[11-14] However, the low electron mobility and poor thermal conductivity of $Ga_2O_3$ may limit device performance. Moreover, these approaches face lattice mismatch issues that degrade interfacial quality and hinder further device development.[6,15,16]

Gallium nitride is a more suitable n-type semiconductor to complement p-type diamond in forming heterojunction diodes. Semiconductor grafting provides a viable approach for creating diamond/GaN p–n heterojunctions.[17-31] This technique enables the integration of lattice-mismatched single-crystalline semiconductors, mitigating lattice and thermal expansion mismatches associated with conventional epitaxial growth.[20,22-24,28,31] Recently, initial p-diamond/n-GaN diodes employing ALD-deposited aluminum oxide as an interfacial layer have been demonstrated, highlighting the effectiveness of the grafting approach.[31] However, as diamond/GaN grafted heterostructures remain at an early stage of development, understanding the band alignment in grafted diamond/GaN p–n junctions is essential for optimizing device performance.

X-ray photoelectron spectroscopy (XPS) is a well-established technique for probing interfacial chemistry and determining band alignment.[17,18,21-25,28] In this work, we investigate diamond/$Al_2O_3$/GaN and diamond/$Al_2O_3$/$SiN_x$/GaN heterojunctions using XPS. The band alignment is constructed by determining the valence-band offset (VBO) and conduction-band offset (CBO) between p-diamond and n-GaN using the Kraut method.[17,21,32] The extracted positive electron affinity of diamond suggests oxygen-related bonding at the diamond/$Al_2O_3$ interface.[33,34] Compared with the diamond/$Al_2O_3$/GaN heterostructure, the diamond/$Al_2O_3$/$SiN_x$/GaN heterostructure exhibits larger band offsets. This modification is likely associated with a change in the interfacial electrostatic potential, possibly resulting from a reduced density of positive fixed charges in the interfacial dielectric near the diamond/$Al_2O_3$ interface after insertion of the ultrathin $SiN_x$ interlayer. These results demonstrate that the band structure of grafted p-diamond/n-GaN heterostructures can be tuned through interfacial-layer engineering.

## II. EXPERIMENT SECTION

The fabrication of grafted diamond/GaN heterostructures consists of three stages: preparation of a single-crystalline diamond nanomembrane (SCD NM), preparation of the GaN host substrate with interfacial layers, and formation of the grafted p-diamond/n-GaN heterostructures, as illustrated in **Fig. 1**.

The SCD NM was obtained from a 5.7 μm-thick boron-doped p-type diamond layer epitaxially grown on a nominally (100)-oriented intrinsic high-pressure high-temperature (HPHT) diamond substrate by microwave plasma chemical vapor deposition. The diamond surface was chemically–mechanically polished to achieve sub-nanometer roughness [**Fig. 1(a1)**]. Details of the diamond growth and polishing processes are reported elsewhere.[31,35,36] Hydrogen ion implantation was then carried out at 225 keV with a dose of $2 \times 10^{17}$ $cm^{-2}$, forming a hydrogen-rich layer approximately 1 μm beneath the diamond surface [**Fig. 1(a2)**]. Electrochemical etching (EC) in deionized (DI) water under an applied voltage of 80 V selectively removed the implanted defect-rich layer, resulting in the release of a freestanding SCD NM [**Fig. 1(a3)**]. The released membrane was subsequently retrieved using a polydimethylsiloxane (PDMS) stamp [**Fig. 1(a4) and (a5)**].

For the GaN host substrate, n-/$n^+$-GaN layers were epitaxially grown on AlN/SiC templates by metal–organic chemical vapor deposition, consisting of a 90 nm AlN buffer layer, a 155 nm Si-doped n-type GaN layer ($6 \times 10^{19}$ $cm^{-3}$), and a 180 nm unintentionally doped GaN layer [**Fig. 1(b)**]. Prior to interfacial layer deposition, the GaN surface was cleaned sequentially with acetone, isopropyl alcohol, and DI water, followed by piranha treatment and diluted hydrochloric acid and hydrofluoric acid etching to remove organic contaminants and native oxides.[37,38] Interfacial layers were then deposited by atomic layer deposition using an Ultratech/Cambridge Nanotech Savannah S200 system. For the first configuration, five cycles of $Al_2O_3$ (0.1 nm/cycle) were deposited using trimethylaluminum and $H_2O$ at 200 °C to form an $Al_2O_3$/GaN structure [**Fig. 1(b1)**]. For the second configuration, 33 cycles of $SiN_x$ (0.015 nm/cycle) were first deposited using tris(dimethylamino)silane at 300 °C, followed by 4 cycles of $Al_2O_3$, forming an $Al_2O_3$/$SiN_x$/GaN structure [**Fig. 1(b2)**]. The total interfacial layer thickness in both cases was approximately 0.5 nm.

Finally, the SCD NMs supported by PDMS were transferred onto the prepared GaN substrates [**Fig. 1(c1)**] and [**Fig. 1(c2)**]. The grafted diamond/$Al_2O_3$/GaN and diamond/$Al_2O_3$/$SiN_x$/GaN heterostructures were formed by thermal annealing at 200 °C for 1 h, with ramping and cooling durations of 1 h and 1.5 h, respectively. This annealing condition was optimized to promote strong interfacial bonding while accommodating the thermal expansion mismatch between diamond and GaN. Optical images of the fabricated heterostructures are provided in **Fig. S1** in the **Supplementary Material**.

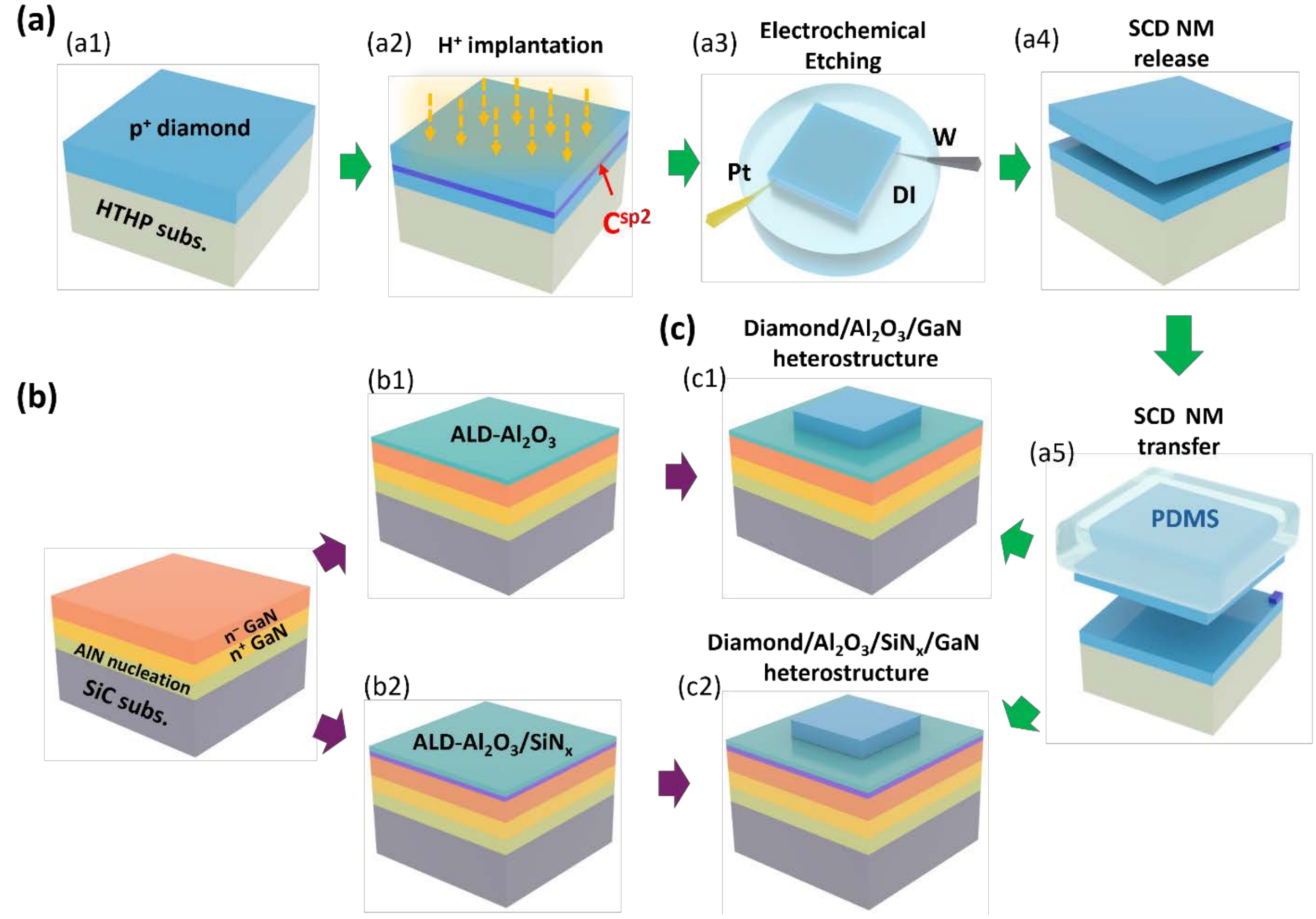

***Fig. 1.*** *Process flow for the fabrication of grafted diamond/GaN heterostructures used for band-alignment studies. (a) Release of the SCD NM by hydrogen ion implantation followed by EC. In panel (a3), Pt and W denote platinum and tungsten needles used to supply the electrical current.(b) Preparation of n-GaN host substrates with (b1)* $Al_2O_3$ *and (b2)* $Al_2O_3/SiN_x$ *interfacial layers, followed by formation of the grafted (c1) diamond/*$Al_2O_3$*/GaN and (c2) diamond/*$Al_2O_3/SiN_x$*/GaN heterostructures.*

Prior to XPS characterization, the SCD NMs in the two p-diamond/n-GaN heterostructures were thinned from their original thickness of ~950 nm to ≤10 nm using a controlled plasma etching process to match the XPS sampling depth (~10 nm). The diamond layers were etched by inductively coupled plasma reactive ion etching (ICP-RIE) using an $O_2/CF_4$ gas mixture (8/2 sccm) at 20 mTorr, with an RIE power of 80 W and ICP power of 800 W, yielding an etch rate of ~2 nm/s.

To access the diamond/GaN interface without over-etching, the diamond was first thinned to ~50 nm based on the calibrated etch rate, followed by stepwise etching in 5 s increments. After each step, XPS measurements were performed using a Thermo Scientific K-Alpha spectrometer with an Al Kα source (hν = 1486.68 eV) to monitor the etching progress. The interface was identified by the simultaneous appearance of strong C 1s and Ga 3d core-level signals from diamond and GaN, respectively. Details of the thinning procedure for both the diamond/$Al_2O_3$/GaN and diamond/$SiN_x$/$Al_2O_3$/GaN heterostructures are provided in **Fig. S2** in the **Supplementary Material**.

## III. RESULTS AND DISCUSSION

To obtain the information required to construct the band alignment at the p-diamond/n-GaN heterointerface by the Kraut method,[17,21,32] five samples were prepared for XPS analysis, as illustrated in **Fig. 2**. A bare n-GaN sample [**Fig. 2(a)**] was prepared using the same cleaning procedure employed for the grafted structures, as described in the Experimental section. The Ga 3d core-level and valence-band maximum (VBM) spectra collected from the bare n-GaN sample served as reference data for constructing the band alignment of both the grafted diamond/$Al_2O_3$/GaN and diamond/$Al_2O_3$/$SiN_x$/GaN heterostructures.

The C 1s core-level and VBM spectra of bare diamond were obtained from the SCD NMs of the as-prepared diamond/$Al_2O_3$/GaN and diamond/$Al_2O_3$/$SiN_x$/GaN heterostructures, as shown in **Figs. 2(b1)** and **2(b2)**, prior to the thinning process. Subsequently, interface samples of the grafted

diamond/$Al_2O_3$/GaN and diamond/$Al_2O_3$/$SiN_x$/GaN heterostructures were prepared for XPS investigation after thinning, as illustrated in **Figs. 2(c1)** and **2(c2)**.

The VBO between p-diamond and n-GaN can be determined using the method proposed by Kraut et al.,[17,21,32]

$$\Delta E_v = \left(E_{C\,1s}^{Diamond} - E_{VBM}^{Diamond}\right) - \left(E_{Ga\,3d}^{GaN} - E_{VBM}^{GaN}\right) + \left(E_{Ga\,3d}^{GaN} - E_{C\,1s}^{Diamond}\right)_{interface}, \quad (1)$$

where $\Delta E_v$ is the VBO, $E_{C\,1s}^{Diamond} - E_{VBM}^{Diamond}$ represents the binding-energy difference between the C 1s core-level peak and the VBM of p-diamond, $E_{Ga\,3d}^{GaN} - E_{VBM}^{GaN}$ is the binding-energy difference between the Ga 3d core-level peak and the VBM of n-GaN, and $\left(E_{Ga\,3d}^{GaN} - E_{C\,1s}^{Diamond}\right)_{interface}$ denotes the binding-energy difference between the Ga 3d and C 1s core-level peaks measured at the p-diamond/n-GaN interface.

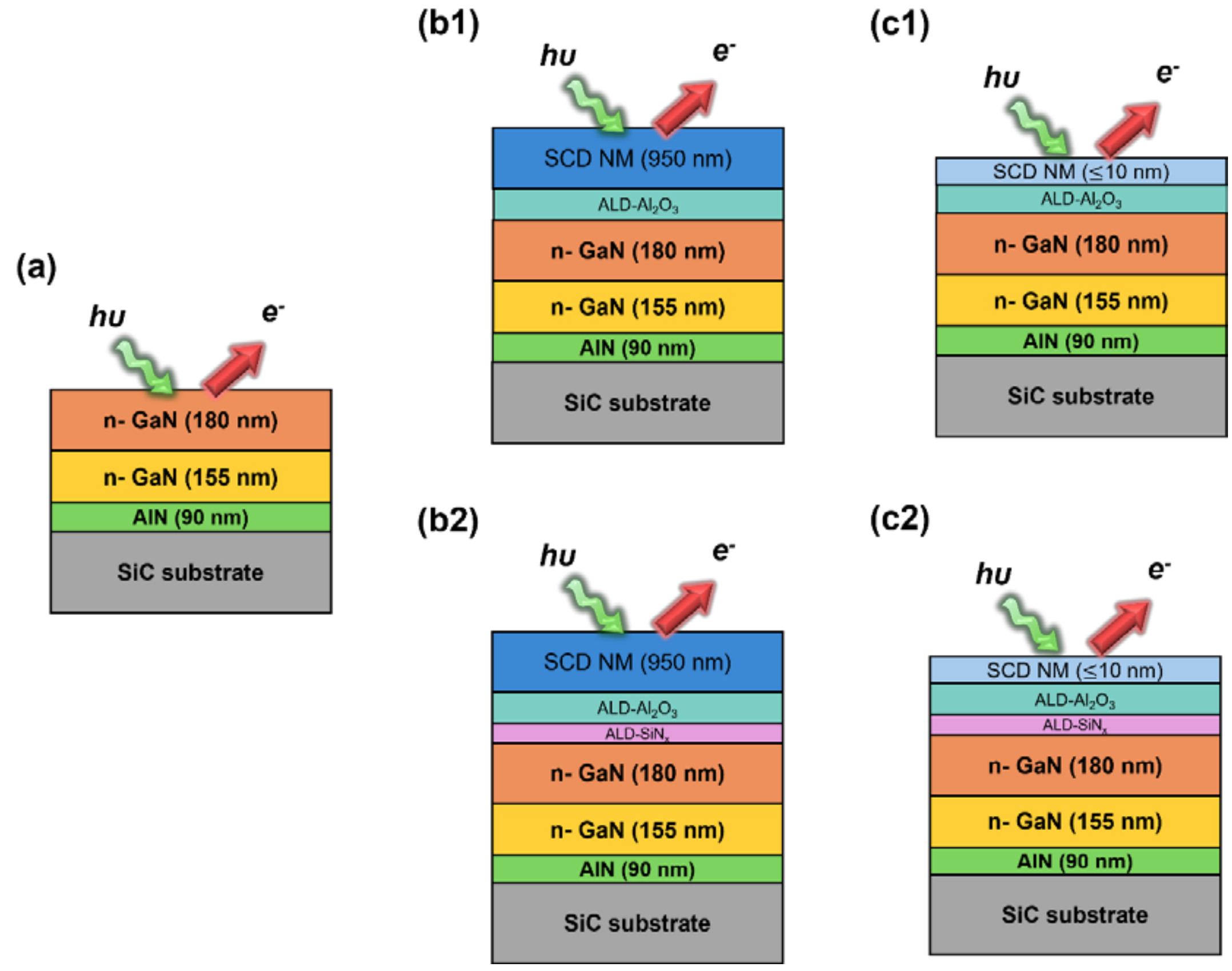

***Fig. 2.*** *Prepared samples used for XPS characterization: (a) bare n-GaN reference sample; heterostructure samples of (b1) diamond/$Al_2O_3$/GaN and (b2) diamond/$Al_2O_3$/$SiN_x$/GaN; and interface samples of (c1) grafted diamond/$Al_2O_3$/GaN and (c2) diamond/$Al_2O_3$/$SiN_x$/GaN.*

**Figure 3** presents the XPS results used to construct the band alignment of the grafted diamond/$Al_2O_3$/GaN heterostructure. For the bare n-GaN sample, the Ga 3d core-level spectrum is shown in **Fig. 3(a1)**. The deconvoluted spectrum exhibits a dominant component attributed to Ga–N bonding centered at 19.95 eV. The VBM of the bare n-GaN sample was determined to be 2.30 eV, as shown in **Fig. 3(a2)**. These XPS-derived values were also used as reference data for the subsequent band-alignment analysis of the diamond/$Al_2O_3$/$SiN_x$/GaN heterostructure.

For the SCD NM of the as-prepared diamond/$Al_2O_3$/GaN heterostructure, the C 1s core-level spectrum is shown in **Fig. 3(b1)**. The deconvoluted spectrum exhibits a pronounced $sp^3$-related C–C bonding component centered at 285.17 eV. The VBM of the SCD NM was determined to be 1.55 eV, as shown in **Fig. 3(b2)**. In addition, a strong $sp^2$-related C=C bonding component is observed at 284.25 eV. The energy separation between the $sp^3$ and $sp^2$ components is 0.92 eV, consistent with previously reported reproducible values ranging from 0.5 to 1.1 eV.[39-41] The pronounced $sp^2$-related surface component is likely associated with the hydrogen ion implantation process used during the SCD NM release, which introduces unsaturated carbon atoms and dangling bonds in the near-surface region.[42,43]

For the grafted diamond/$Al_2O_3$/GaN interface sample, the Ga 3d and C 1s core-level spectra are shown in **Fig. 3(c1)** and **Fig. 3(c2)**, respectively. In the deconvoluted Ga 3d spectrum, a dominant Ga-N component located at 20.67 eV and a secondary Ga-O component at 21.88 eV are observed. Compared with the bare n-GaN sample, the Ga 3d spectrum of the interface sample appears noticeably broadened, which is attributed to the increased composition of the Ga–O component. This observation suggests enhanced interfacial interaction, including Ga-O-related bonding, between $Al_2O_3$ and GaN after thermal annealing.[43,44]

In the deconvoluted C 1s spectrum, a dominant $sp^3$-related C–C bonding component is observed at 286.01 eV, accompanied by two weaker components located at 285.31 eV and 287.14 eV, which are attributed to $sp^2$-related C=C bonding and C–O bonding, respectively. Compared with the surface of the SCD NM, the reduced $sp^2$-related component observed for the grafted diamond/$Al_2O_3$/GaN interface sample may be attributed to the relatively lower degree of implantation-induced damage on the backside of the released SCD NM and/or the formation of C–O bonding at the diamond/$Al_2O_3$ interface, as evidenced by the presence of a weak C–O peak. In this band-alignment analysis, the binding energies of the Ga–N and C–C components in the deconvoluted Ga 3d and C 1s spectra were used for the VBO calculation, as these correspond to the dominant chemical bonding states in GaN and diamond, respectively. The parameters used for the band-alignment calculation of the grafted diamond/$Al_2O_3$/GaN heterostructure are summarized in **Table I**, together with the fitted full width at half maximum (FWHM) values of all peaks.

By inserting the XPS-derived parameters into **Eq. (1)**, the VBO of the grafted diamond/$Al_2O_3$/GaN heterostructure was determined to be 0.63 eV. Combining this value with the experimental bandgap energies of GaN (3.4 eV)[44] and diamond (5.47 eV)[45], the conduction-band offset (CBO) was calculated to be 2.70 eV using

$$\Delta E_c = E_g^{GaN} - E_g^{diamond} - \Delta E_v \quad (2)$$

Assuming an ideal grafted diamond/$Al_2O_3$/GaN interface without interface dipoles or surface band bending, the CBO can alternatively be expressed in terms of the electron-affinity difference between GaN and diamond[46]:

$$\Delta E_c = \chi_{GaN} - \chi_{diamond}, \qquad (3)$$

where $\chi_{GaN}$ and $\chi_{diamond}$ denote the electron affinities of GaN and diamond, respectively. Using the CBO extracted from XPS and the reported electron affinity of GaN (4.1 eV)[47], the effective electron affinity of diamond is estimated to be ~1.4 eV. This value is consistent with previously reported electron affinities for oxygen-related diamond surfaces.[33,48] Such a positive electron affinity is likely associated with oxygen-related interfacial bonding at the diamond/$Al_2O_3$ interface rather than a pristine surface termination. This interpretation is supported by the emergence of a weak C–O component in the C 1s spectrum of the grafted diamond/$Al_2O_3$/GaN interface sample [**Fig. 3(c2)**], indicating the presence of oxygen-induced interfacial bonding (e.g., C–O and C–O–Al).

***Table I.*** *Summary of the extracted parameters used for band alignment calculations and analysis of electrostatic potential modulation, including peak center positions and FWHM. VBM denotes the valence band maximum.* $\Delta BE_{Ga\text{-}N}$ *and* $\Delta BE_{C\text{-}C}$ *represent the binding energy shifts of the Ga–N and C–C components at the interface relative to their respective reference samples.*

| **Sample** | ***VBM*** **(eV)** | **Ga 3d** | | | | **C 1s** | | | | | | $\Delta BE_{Ga\text{-}N}$ **(eV)** | $\Delta BE_{C-C}$ **(eV)** |
|---|---|---|---|---|---|---|---|---|---|---|---|---|---|
| | | **Ga-N** | | **Ga-O** | | **C-C** | | **C=C** | | **C-O** | | | |
| | | **Center (eV)** | **FWHM (eV)** | **Center (eV)** | **FWHM (eV)** | **Center (eV)** | **FWHM (eV)** | **Center (eV)** | **FWHM (eV)** | **Center (eV)** | **FWHM (eV)** | | |
| Bare n-GaN | 2.3 | 19.95 | 1.34 | 20.76 | 1.47 | | | | | | | | |
| SCD (on $Al_2O_3$/GaN) | 1.55 | | | | | 285.17 | 1.09 | 284.25 | 1.32 | 286.30 | 1.05 | | |
| *Interface* (Diamond/$Al_2O_3$/GaN) | | 20.67 | 1.39 | 21.88 | 1.69 | 286.01 | 0.90 | 285.31 | 1.51 | 287.14 | 1.27 | 0.72 | 0.84 |
| SCD (on $Al_2O_3$/$SiN_x$/GaN) | 1.65 | | | | | 285.24 | 1.04 | 284.31 | 1.49 | 286.38 | 0.91 | | |
| *Interface* (Diamond/$Al_2O_3$/$SiN_x$/GaN) | | 20.63 | 1.34 | 21.84 | 1.99 | 285.52 | 1.17 | 284.82 | 1.68 | 286.87 | 1.09 | 0.68 | 0.28 |

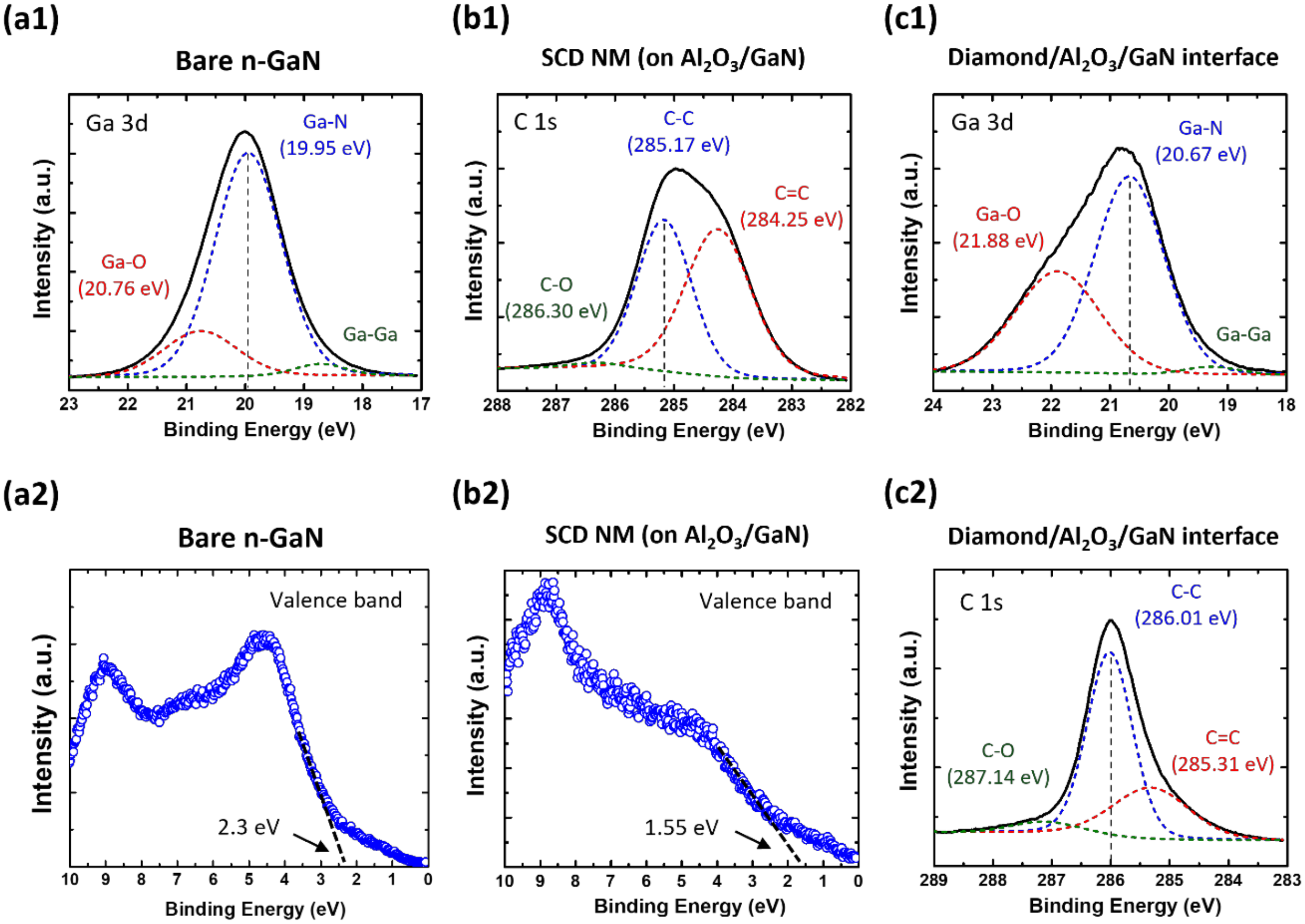

***Fig. 3.*** *XPS spectra used for band alignment analysis. (a1) Ga 3d XPS spectrum of the bare n-GaN reference sample with deconvoluted components. (a2) Valence-band XPS spectrum of the bare n-GaN reference sample with extraction of the VBM value. (b1) C 1s XPS spectrum of the SCD NM from the grafted diamond/$Al_2O_3$/GaN heterostructure with deconvoluted components. (b2) Valence-band XPS spectrum of the SCD NM from the grafted diamond/$Al_2O_3$/GaN heterostructure with extraction of the VBM value. (c1) Ga 3d and (c2) C 1s XPS spectra of the grafted diamond/$Al_2O_3$/GaN interface sample with deconvoluted components.*

In addition to the diamond/$Al_2O_3$/GaN heterostructure, a diamond/$Al_2O_3$/$SiN_x$/GaN heterostructure was also fabricated and characterized by XPS to investigate the influence of the inserted $SiN_x$ interlayer on the interfacial band alignment. The fabrication followed the same procedure as that used for the diamond/$Al_2O_3$/GaN structure, with the only modification being the introduction of an additional $SiN_x$ layer [**Fig. 1(b)**], which serves as an effective passivation layer for the GaN surface.[49–51] To construct the band alignment, the corresponding XPS spectra are presented in **Fig. 4**, together with the reference spectra obtained from the bare n-GaN sample shown in **Fig. 3(a1)**. **Figure 4(a1)** shows the C 1s core-level spectrum of the SCD NM in the as-prepared diamond/$Al_2O_3$/$SiN_x$/GaN heterostructure. The deconvoluted spectrum is dominated by a pronounced $sp^3$-related C–C component centered at 285.24 eV, accompanied by a weaker $sp^2$-related C=C component at 284.31 eV. The VBM of the SCD NM was determined to be 1.65 eV, as shown in **Fig. 4(a2)**. Notably, the energy separation between the $sp^3$-related C 1s core-level peak and the VBM of the SCD NM in the diamond/$Al_2O_3$/$SiN_x$/GaN heterostructure is 283.59 eV, which is very close to the corresponding value of 283.62 eV obtained for the SCD NM of the diamond/$Al_2O_3$/GaN heterostructure [**Fig. 3(b)**]. The similar C 1s core-level–to–VBM energy separations, together with the comparable line shapes of the deconvoluted C 1s spectra and the consistent energy separation between the $sp^3$ and $sp^2$ components, indicate similar near-surface electronic structures and comparable surface electronic quality of the SCD NMs in the two heterostructures.

For the grafted diamond/$Al_2O_3$/$SiN_x$/GaN interface sample, the Ga 3d and C 1s core-level spectra are presented in **Fig. 4(b1)** and **Fig. 4(b2)**, respectively. The deconvoluted Ga 3d spectrum reveals a dominant Ga–N component centered at 20.63 eV, together with a secondary Ga–O component located at 21.84 eV. In the deconvoluted C 1s spectrum, the $sp^3$-related C–C bonding component dominates at 285.52 eV, accompanied by two weaker features at 284.82 eV and 286.87 eV, which are assigned to $sp^2$-related C=C bonding and C–O bonding, respectively.

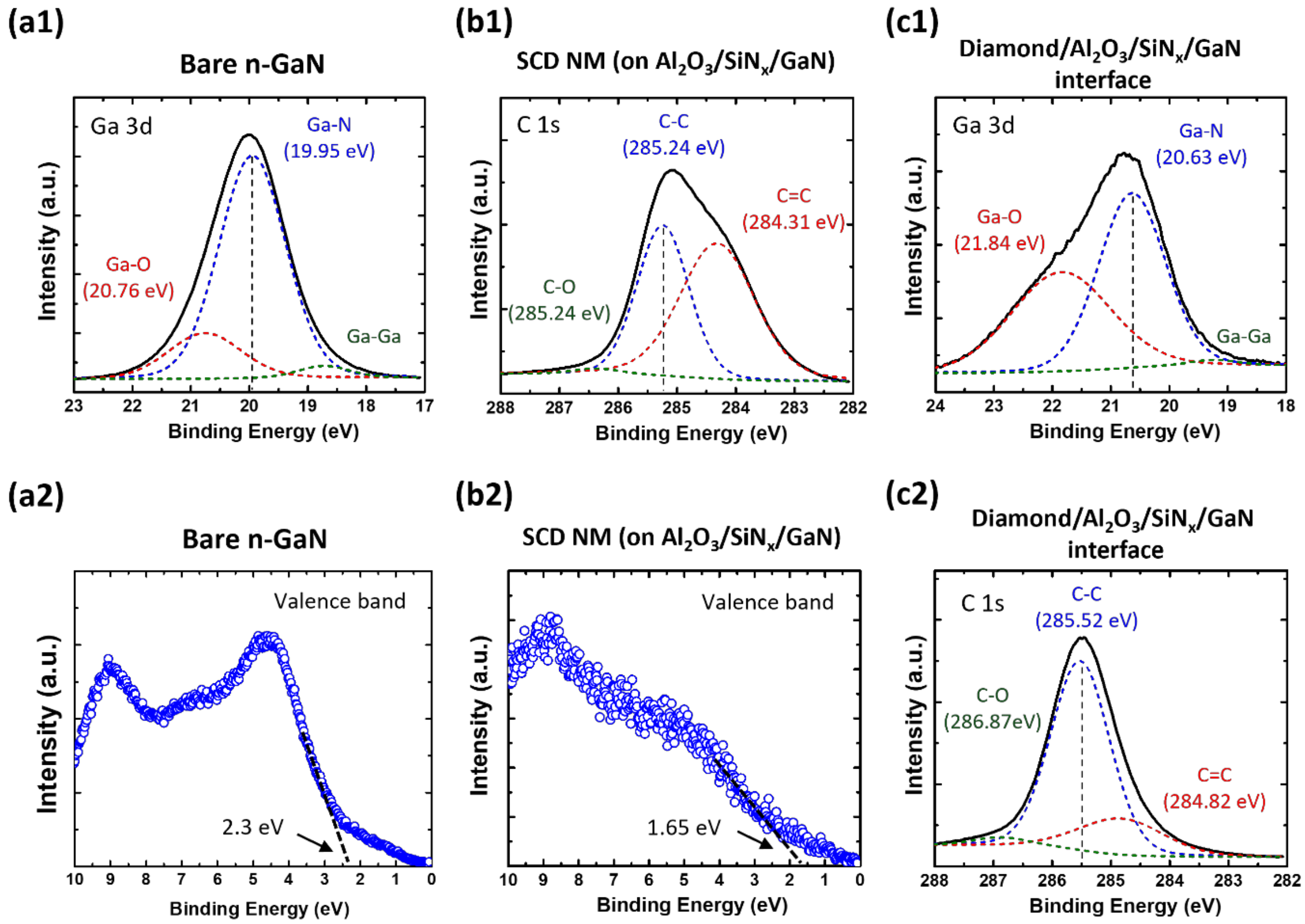

***Fig. 4.*** *XPS spectra used for band alignment analysis. (a1) Ga 3d XPS spectrum of the bare n-GaN reference sample with deconvoluted components. (a2) Valence-band XPS spectrum of the bare n-GaN reference sample with extraction of the VBM value. (b1) C 1s XPS spectrum of the SCD NM from the grafted diamond/*$Al_2O_3/SiN_x/GaN$ *heterostructure with deconvoluted components. (b2) Valence-band XPS spectrum of the SCD NM from the grafted diamond/*$Al_2O_3/SiN_x/GaN$ *heterostructure with extraction of the VBM value. (c1) Ga 3d and (c2) C 1s XPS spectra of the grafted diamond/*$Al_2O_3/SiN_x/GaN$ *interface sample with deconvoluted components.*

The parameters employed for constructing the band alignment of the diamond/$Al_2O_3/SiN_x/GaN$ heterostructure are also summarized in **Table I**, together with those of the diamond/$Al_2O_3$/GaN heterostructure for direct comparison. By substituting the XPS-derived parameters into **Eq. (1)**, the VBO of the diamond/$Al_2O_3/SiN_x/GaN$ heterostructure was determined to be 1.05 eV. Using **Eq. (2)** and the experimental bandgap values of GaN and diamond, the corresponding CBO was calculated to be 3.12 eV.

The band-alignment diagrams constructed from the XPS measurements for the diamond/$Al_2O_3$/GaN and diamond/$Al_2O_3$/$SiN_x$/GaN heterostructures are shown in **Figs. 5(a)** and **5(b)**, respectively. Both heterostructures exhibit type-II band alignment. Upon insertion of the ultrathin $SiN_x$ interlayer prior to $Al_2O_3$ deposition, the VBO increases from 0.63 to 1.05 eV, while the CBO increases from 2.70 to 3.12 eV. The substantial 0.42 eV increase in both band offsets highlights the effectiveness of the inserted $SiN_x$ layer in tailoring the interfacial band alignment of the grafted diamond/GaN heterojunction. The observed modification in band offsets is attributed to a change in the interfacial electrostatic potential. Such a change may originate from interfacial dipoles associated with chemical bonding and charge redistribution, as well as from fixed charges within the interfacial dielectric stack. These effects can modify the vacuum-level alignment and the band bending near the interface, thereby leading to the observed variations in both VBO and CBO.[52–54]

To elucidate the origin of the pronounced difference in band offsets between the diamond/$Al_2O_3$/GaN and diamond/$Al_2O_3$/$SiN_x$/GaN heterostructures, the Ga 3d and C 1s core levels of all samples were systematically analyzed. Key parameters, including the FWHM of each component and the core-level binding energy shifts (ΔBE) between the reference samples and the grafted diamond/GaN heterostructures, are summarized in **Table I**, where FWHM and ΔBE provide insight into the extent of charge-related effects and variations in the interfacial electrostatic potential. Compared with the bare n-GaN reference sample, the binding energies of the Ga–N and Ga–O components in the Ga 3d spectrum of the diamond/$Al_2O_3$/GaN interface shift from 19.95 to 20.67 eV ($\Delta BE_{Ga\text{-}N}$ = 0.72 eV) and from 20.76 to 21.88 eV ($\Delta BE_{Ga\text{-}O}$ = 1.12 eV), respectively, as shown in **Figs. 3(a1)** and **3(c1)**. Only negligible changes in FWHM are observed for both components, suggesting that charging-induced peak distortion is minimal. The shift of the Ga–N component is primarily attributed to a change in the electrostatic potential near the $Al_2O_3$/GaN interface. In contrast, the additional 0.40 eV shift ($\Delta BE_{Ga\text{-}O}$ - $\Delta BE_{Ga\text{-}N}$) of the Ga–O component, together with the increased Ga–O/Ga–N intensity ratio, indicates a modified chemical environment, likely associated with the formation of $GaO_x$-related bonding on GaN during $Al_2O_3$ deposition.[55]

For the C 1s spectrum of the diamond/$Al_2O_3$/GaN interface sample, the binding energies of the C–C and C=C components shift from 285.17 to 286.01 eV ($\Delta BE_{C-C}$ = 0.84 eV) and from 284.25 to 285.31 eV ($\Delta BE_{C=C}$ = 1.06 eV), respectively, relative to the corresponding SCD reference sample, as shown in **Figs. 3(b1)** and **3(c2)**. The shift of the C–C component, together with the negligible change in FWHM, indicates a substantial electrostatic-potential change on the diamond side. The additional 0.22 eV shift ($\Delta BE_{C=C}$ - $\Delta BE_{C\text{-}C}$) of the $sp^2$-related component relative to the $sp^3$-related component suggests a modified local chemical environment, potentially associated with oxygen-related bonding at the diamond/$Al_2O_3$ interface.[56]

For the diamond/$Al_2O_3$/$SiN_x$/GaN interface sample, the Ga–N and Ga–O components in the Ga 3d spectrum shift from 19.95 to 20.63 eV ($\Delta BE_{Ga\text{-}N}$ = 0.68 eV) and from 20.76 to 21.84 eV ($\Delta BE_{Ga\text{-}O}$ = 1.08 eV), respectively, relative to the bare n-GaN reference, as shown in **Figs. 4(a1)** and **4(c1)**. Similarly, the C-C and C=C components in the C 1s spectrum shift from 285.24 eV to 285.52 eV ($\Delta BE_{C-C}$ = 0.28 eV) and from 284.31 eV to 284.82 eV ($\Delta BE_{C=C}$ = 0.51 eV), respectively, relative to the corresponding SCD reference sample, as shown in **Figs. 4(b1)** and 4**(c2)**.

It is worth noting that the relative shifts between the Ga–O and Ga–N components ($\Delta BE_{Ga\text{-}O}$ - $\Delta BE_{Ga\text{-}N}$) are ~0.4 eV for both the diamond/$Al_2O_3$/GaN and diamond/$Al_2O_3$/$SiN_x$/GaN heterostructures. Similarly, the differences between the C=C and C–C components ($\Delta BE_{C=C}$ - $\Delta BE_{C\text{-}C}$) are 0.22 and 0.23 eV for the diamond/$Al_2O_3$/GaN and diamond/$Al_2O_3$/$SiN_x$/GaN heterostructures, respectively. The close agreement in these values suggests that the local chemical environments at the interfaces remain largely comparable, despite the insertion of the ultrathin $SiN_x$ interlayer. This apparently unchanged chemical state on the insulator/GaN side may be attributed to the extremely small thickness of the inserted $SiN_x$ layer (~0.1 nm), which is much smaller than the XPS information depth (~3–5 nm). As a result, the direct contribution of the $SiN_x$ layer to the measured XPS signal is expected to be limited.[57]

Subsequently, the variations in the Ga 3d and C 1s core-level energies of n-GaN and p-diamond in the two heterostructures were further analyzed by rewriting **Eq. (1)** into **Eq. (4)**, as shown below.

$$\Delta E_v = \left(E_{C\,1s}^{bulk} - E_{C\,1s}^{interface}\right)_{\text{diamond}} - \left(E_{Ga\,3d}^{bulk} - E_{Ga\,3d}^{interface}\right)_{\text{GaN}} - \left(E_{VBM}^{Diamond} - E_{VBM}^{GaN}\right), \quad (4)$$

where $E_{C\,1s}^{bulk} - E_{C\,1s}^{interface}$ represents the energy difference between the C 1s core-level positions measured from the bare SCD reference sample and the corresponding interface sample, and $E_{Ga\,3d}^{bulk} - E_{Ga\,3d}^{interface}$ denotes the corresponding energy difference for the Ga 3d core level between the bare n-GaN reference and the interface sample. The $E_{Ga\,3d}^{bulk} - E_{Ga\,3d}^{interface}$ value ($\Delta BE_{\text{Ga-N}}$) in the diamond/$Al_2O_3$/GaN heterostructure (0.72 eV) is nearly identical to that in the diamond/$Al_2O_3$/$SiN_x$/GaN heterostructure (0.68 eV). This similarity indicates that the electrostatic potential on the n-GaN side remains largely unchanged despite the insertion of the ultrathin $SiN_x$ layer. In contrast, the $E_{C\,1s}^{bulk} - E_{C\,1s}^{interface}$ values ($\Delta BE_{\text{C-C}}$) are 0.84 eV and 0.28 eV for the diamond/$Al_2O_3$/GaN and diamond/$Al_2O_3$/$SiN_x$/GaN heterostructures, respectively. The substantial difference of 0.56 eV indicates that the dominant electrostatic-potential change occurs on the diamond side. This difference consequently leads to noticeable variations in the calculated VBO and CBO in the XPS-derived band alignments, as shown in **Fig. 5(a)** and **Fig. 5(b)**.

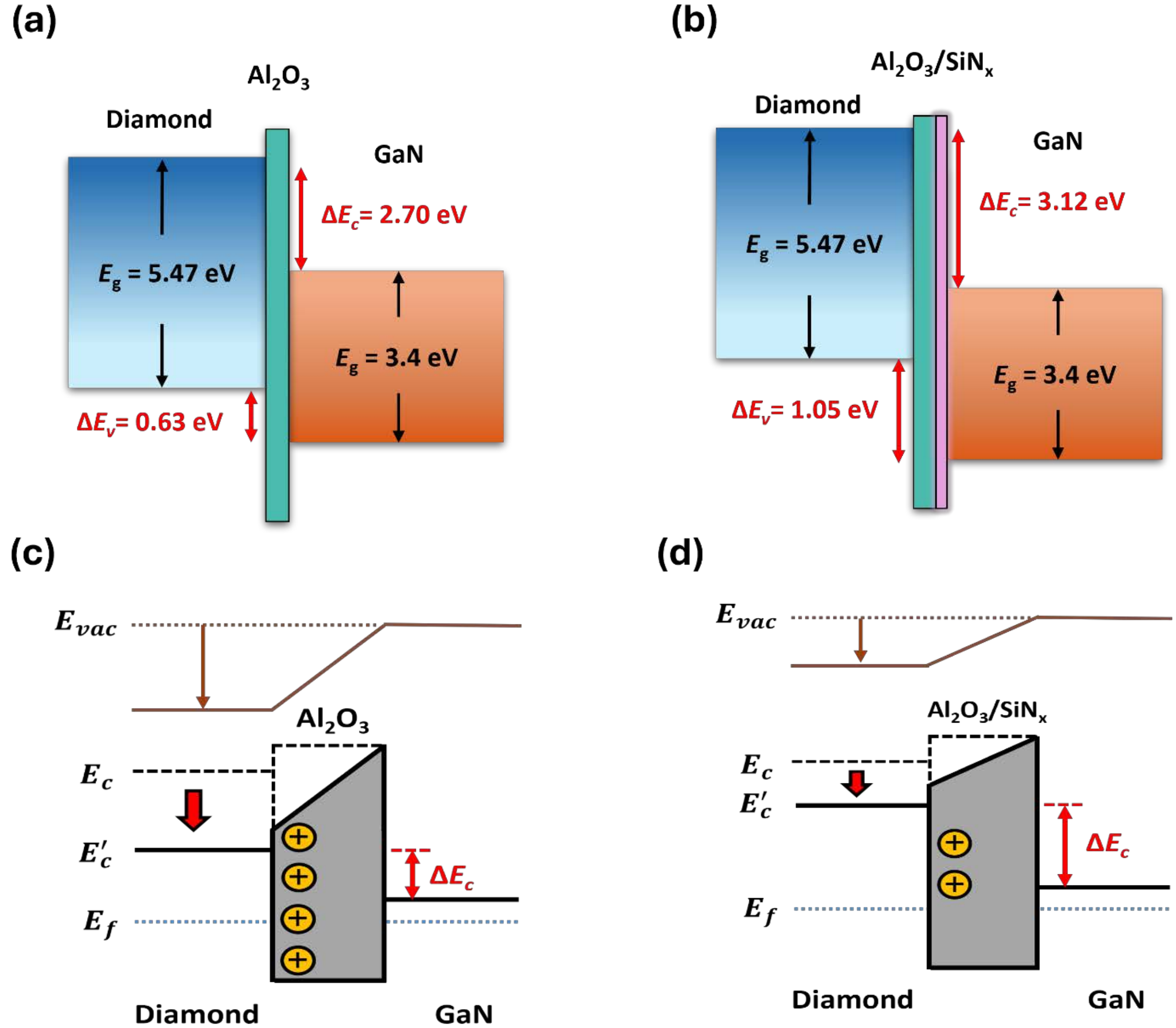

***Fig. 5.*** *Band-alignment diagrams of the grafted (a) diamond/$Al_2O_3$/GaN and (b) diamond/$Al_2O_3$/$SiN_x$/GaN heterostructures constructed from XPS measurements. (c, d) Schematic band diagrams illustrating the proposed modulation of the electrostatic potential by positive fixed charges in the interfacial dielectric near the diamond/$Al_2O_3$ interface for the corresponding heterostructures.*

From these observations, the chemical environments on both the p-diamond and n-GaN sides remain nearly unchanged for the diamond/$Al_2O_3$/GaN and diamond/$Al_2O_3$/$SiN_x$/GaN heterostructures. Meanwhile, the Ga 3d core-level shifts on the n-GaN side are nearly identical for the two samples, whereas significantly larger C 1s core-level shifts are observed on the p-diamond side. These results indicate that the substantial change in electrostatic potential occurs primarily on the p-diamond side.

One possible explanation is the formation of an interfacial dipole induced by the insertion of the ultrathin $SiN_x$ layer, which could introduce an electrostatic potential change at the interface.[51] However, the screening effects in p-diamond and n-GaN are expected to be of similar magnitude due to their comparable p- and n-type doping concentrations.[58] Under such conditions, an interfacial dipole would be expected to induce electrostatic potential changes on both sides of the junction simultaneously, which is inconsistent with the experimental observations.

A more plausible explanation is that the density of positive fixed charges in the interfacial dielectric, located at or near the diamond/$Al_2O_3$ interface, is reduced after insertion of the ultrathin $SiN_x$ layer. Deposition of $SiN_x$ prior to $Al_2O_3$ may modify the nucleation behavior and defect chemistry of the subsequently deposited $Al_2O_3$ film.[59] A reduction in the density of these positive fixed charges would decrease the electric field across the interfacial dielectric stack and thereby reduce the electrostatic-potential drop and band bending on the adjacent diamond side.[54] This interpretation is consistent with the smaller C 1s core-level shift ($\Delta BE_{C\text{-}C}$) observed in the diamond/$Al_2O_3$/$SiN_x$/GaN heterostructure.

To illustrate this mechanism, schematic band diagrams of the diamond/$Al_2O_3$/GaN and diamond/$Al_2O_3$/$SiN_x$/GaN heterostructures are presented in **Figs. 5(c)** and **5(d)**, respectively. In the diamond/$Al_2O_3$/GaN heterostructure [**Fig. 5(c)**], positive fixed charges in the interfacial dielectric near the diamond/$Al_2O_3$ interface induce an electric field across the dielectric, resulting in a larger electrostatic-potential drop and stronger band bending on the diamond side. After insertion of the ultrathin $SiN_x$ layer [**Fig. 5(d)**], the density of these positive fixed charges is reduced, thereby weakening the electrostatic-potential drop on the diamond side. As a result, the diamond bands shift to higher energy relative to the Fermi level, leading to larger band offsets at the diamond/GaN heterointerface. This scenario is consistent with the larger VBO and CBO values extracted from the XPS measurements of the diamond/$Al_2O_3$/$SiN_x$/GaN heterostructure.

The increased VBO and CBO at the diamond/GaN interface resulting from the insertion of the ultrathin $SiN_x$ interlayer provide higher carrier barriers for both electrons and holes. Moreover, the $SiN_x$ interlayer may passivate interfacial defect states, thereby reducing interface recombination and suppressing reverse leakage current. These combined effects are beneficial for improving the rectification characteristics of the p-diamond/n-GaN diode.

However, excessively large band offsets may also hinder carrier injection across the heterointerface, potentially increasing the turn-on voltage or limiting the forward current. Therefore, careful engineering of the band alignment through optimization of the $Al_2O_3$/$SiN_x$ interfacial layers is necessary to achieve an optimal balance between leakage suppression and efficient carrier transport. Such interface engineering provides a promising pathway toward high-performance, high-power diamond/GaN p–n heterojunction diodes.

## IV. CONCLUSIONS

In this work, grafted diamond/$Al_2O_3$/GaN and diamond/$Al_2O_3$/$SiN_x$/GaN heterostructures were successfully fabricated and investigated by XPS. Using the Kraut method, the valence-band offset (VBO) and conduction-band offset (CBO) of the diamond/$Al_2O_3$/GaN heterostructure were determined to be 0.63 and 2.70 eV, respectively, while those of the diamond/$Al_2O_3$/$SiN_x$/GaN heterostructure were determined to be 1.05 and 3.12 eV. The introduction of an ultrathin $SiN_x$ interlayer prior to $Al_2O_3$ deposition therefore enables effective tuning of the band alignment at the p-diamond/n-GaN interface, resulting in a band-offset increase of approximately 0.42 eV.

Analysis of the XPS spectra suggests that the observed increase in both VBO and CBO is associated with a modification of the interfacial electrostatic potential. The experimental results are consistent with a reduced density of positive fixed charges in the interfacial dielectric near the diamond/$Al_2O_3$ interface after insertion of the $SiN_x$ interlayer, which weakens the electrostatic-potential drop on the diamond side. Consequently, diamond/$Al_2O_3$/$SiN_x$/GaN p–n heterojunction diodes are expected to exhibit improved rectification behavior compared with diamond/$Al_2O_3$/GaN diodes because of the increased carrier barriers and the possible suppression of interface recombination and reverse leakage current.

Further investigations, including ultraviolet photoelectron spectroscopy (UPS), scanning transmission electron microscopy (STEM), and capacitance–voltage (C–V) measurements, are needed to further clarify the origin of the band-offset modulation. These findings provide useful insight into interfacial band-alignment engineering in grafted diamond/GaN p–n heterojunctions and may facilitate the development of high-power electronic devices based on this material platform.

**Acknowledgment**

The work was supported by the Defense Advanced Research Projects Agency (DARPA) under contract 140D04-24-C-0061. The views, opinions and/or findings expressed are those of the authors and should not be interpreted as representing the official views or policies of DARPA or the U.S. Government.

## DATA AVAILABILITY

The data that supports the findings of this study are available from the corresponding author upon reasonable request.